\newtheorem{lemma}{Lemma}
\newtheorem{theorem}{Theorem}
\newtheorem{remark}{Remark}

\newtheorem{definition}{Definition}

\documentclass[12pt]{article}
\usepackage{amsfonts}

\usepackage{graphics}

\input{tcilatex}

\begin{document}

\title{Quantum Grammars}
\author{V.A.Malyshev \thanks{%
Postal address:INRIA - Domaine de Voluceau, Rocquencourt, BP105 - 78153 - Le
Chesnay Cedex, France.}}
\begin{titlepage}   
\maketitle 

\begin{abstract}
We consider quantum (unitary) continuous time evolution of spins on a
lattice together with quantum evolution of the lattice itself. In physics
such evolution was discussed in connection with quantum gravity. It is also
related to what is called quantum circuits, one of the incarnations of a
quantum computer. We consider simpler models for which one can obtain exact
mathematical results. We prove existence of the dynamics in both
Schroedinger and Heisenberg pictures, construct KMS states on appropriate $%
C^{\ast }$-algebras.

We show (for high temperatures) that for each system where the lattice
undergoes quantum evolution, there is a natural scaling leading to a quantum
spin system on a fixed lattice $Z$, defined by a renormalized Hamiltonian.
\end{abstract}
\end{titlepage}
\pagebreak

\section{\protect\bigskip Introduction}

Practical quantum computation has not yet started but many standard notions
of the computer science have already been generalized, giving rise to the
quantum computer science, see recent reviews \cite{aha, tsi, ste}. Here we
give a definition of a quantum grammar similar to the definition of a random
grammar, given in \cite{m1}.

A very particular case of quantum grammars are quantum spin systems, popular
standard models in statistical physics and quantum field theory. Quantum
grammar can be considered as a quantum spin system on a quantum lattice,
that is the lattice itself is a quantum object subject to a unitary
evolution. It is quite in a spirit of some approaches to the quantum
gravity, where space is quantized, but the time remains classical and
one-dimensional.

Here we consider questions pertinent to physical systems rather than to the
computer science. We show how standard quantum spin systems (spin represents
the matter) on the lattice $Z$ (lattice represents the space) can emerge
from KMS states on the $C^{\ast }$-algebras corresponding to quantum
grammars. The term grammar refers normally to one-dimensional systems.
Higher dimensional objects are called graph grammars in computer science.
Higher dimension means only that it is not one-dimensional. The terms spin
graph, spin complex or spin network are used instead of ''higher dimensional
grammars''.

One of our goals is to show that already in one dimension these models have
sufficiently interesting structure. The evolution of the space is simple
however. There is no topology, only metrics is important: the space can
expand and compress at any point, expanding and compressing being a quantum
process. However there are phenomena which have no analogs in the
statistical physics and quantum field theory living on a classical space.

The correspondence between grammars and quantum grammars are as between
classical and quantum computation. We consider continuous time evolution
which allows the grammars be far from context free. Thus there are no
''no-go'' theorems as for the discrete time, see \cite{mey}. We prove
selfadjointness of the Hamiltonian which gives the unitary evolution on a
Hilbert space and an automorphism group of some hyperfinite $C^{\ast }$%
-algebra. We show that there is a transition in the parameters (the
temperature and the cosmological constant) when the KMS state exists or not.
In the latter case we define renormalised KMS states, the scaling limit of
such renormalized states is a standard quantum spin system.

\section{Symmetric Grammars}

\subsection{Hilbert space and Hamiltonian}

Let $S=\left\{ 1,...,r\right\} $ be a finite set (the alphabet), $L=L(S)$ -
the set of all finite words (including the empty one) $\alpha
=x_{1}...x_{n},x_{i}\in S,$ in this alphabet. Length $n$ of the word $\alpha 
$ is denoted by $\left| \alpha \right| $. Concatenation of two words $\alpha
=x_{1}...x_{n}$ and $\beta =y_{1}...y_{m}$ is defined by 
\[
\alpha \beta =x_{1}...x_{n}y_{1}...y_{m} 
\]
The word $\beta $ is a subword of $\alpha $ if there exist words $\delta $
and $\gamma $ such that $\alpha =\delta \beta \gamma $. Grammar over $S$ is
defined by a finite set $Sub$ of substitutions (productions), that is the
pairs $\delta _{i}\rightarrow \gamma _{i},i=1,...,k=\left| Sub\right|
,\delta _{i},\gamma _{i}\in L$. \ Further on we assume that all $\delta
_{i},\gamma _{i}$ are not empty.

Let $\mathcal{H}=l_{2}(L)$ be the Hilbert space with the orthonormal basis $%
e_{\alpha },\alpha \in L:(e_{\alpha },e_{\beta })=\delta _{\alpha \beta }$
where the function $e_{\alpha }(\beta )=\delta _{\alpha \beta }$. Each
vector $\phi $ of $\mathcal{H}$ a function on the set of words and can be
written as 
\[
\phi =\sum \phi (\alpha )e_{\alpha }\in \mathcal{H},\left\| \phi \right\|
^{2}=\sum \left| \phi (\alpha )\right| ^{2} 
\]
States of the system are wave functions, that is vectors $\phi $ with the
unit norm $\left\| \phi \right\| ^{2}=1$. We shall define dynamics in the
form 
\[
\phi (t)=\exp (itH)\phi (0) 
\]
The Hamiltonian $H$ will be written in terms of operators, which resemble
creation-annihilation operators in quantum field theory. For each $i=1,...,k$
and each integer $j\geq 1$ we define quantum substitutions, that is linear
bounded operators $a_{i}(j)$. If $\alpha =\tau \delta _{i}\rho $ for some
words $\tau ,\rho ,\left| \tau \right| =j-1$, we put 
\[
a_{i}(j)e_{\alpha }=e_{\beta } 
\]
where $\beta =\tau \gamma _{i}\rho $ . Otherwise we put $a_{i}(j)e_{\alpha
}=0$. Adjoint operators $a_{i}^{\ast }(j)$ are defined by 
\[
a_{i}^{\ast }(j)e_{\beta }=e_{\alpha } 
\]
for $\beta =$ $\tau \gamma _{i}\rho $ and $0$ otherwise. Define the formal
Hamiltonian by 
\[
H=\sum_{i=1}^{\left| Sub\right| }\sum_{j=1}^{\infty }(\lambda _{i}a_{i}(j)+%
\overline{\lambda }_{i}a_{i}^{\ast }(j)) 
\]
for some complex $\lambda _{i}$.

We could equally assume that together with the substitution $\delta
_{i}\rightarrow \gamma _{i}$ also its ''inverse'' substitution $\gamma
_{i}\rightarrow \delta _{i}$ belongs to $Sub$. The Hamiltonian then can be
written simply as 
\[
H=\sum_{i=1}^{\left| Sub\right| }\sum_{j=1}^{\infty }\lambda _{i}a_{i}(j) 
\]
We always assume that $H=H^{\ast }$, that is $\lambda _{i}=\overline{\lambda 
}_{j}$ in case $\delta _{j}=\gamma _{i},\gamma _{j}=\delta _{i}$. We shall
use only this representation further on.

$H$ is well-defined and symmetric on the set $D(L)$ of finite linear
combinations of $e_{\alpha }$. These vectors are $C^{\infty }$-vectors for $%
H $, that is $He_{\alpha }\in D(L)$.

\begin{theorem}
$H$ is essentially selfadjoint on $D(L)$.
\end{theorem}

Proof. We shall prove that each vector $\phi \in D(L)$ is an analytic vector
of $H$, that is 
\[
\sum_{k=0}^{\infty }\frac{\left\| H^{k}\phi \right\| }{k!}t^{k}<\infty 
\]
for some $t>0$. It is sufficient to take $\phi =e_{\alpha }$ for some $%
\alpha $. Then the number of pairs $(i,j)$ such that $a_{i}(j)e_{\alpha
}\neq 0$ is not greater than $nk,n=\left| \alpha \right| ,k=\left|
Sub\right| $.

Write the decomposition of $H$ as 
\[
H=\sum_{a}V_{a} 
\]
where $V_{a}$ equals one of $\lambda _{i}a_{i}(j)$. Then 
\begin{equation}
H^{n}e_{\alpha }=\sum_{a_{n},...,a_{1}}V_{a_{n}}...V_{a_{1}}e_{\alpha }=\sum
C_{\beta }e_{\beta }  \label{exp11}
\end{equation}
The maximal length of the words $\beta $ in the expansion of $%
V_{a_{n}}...V_{a_{1}}e_{\alpha }$ does not exceed $\left| \alpha \right|
+C_{1}n,C_{1}=\max (\left| \gamma _{i}\right| -\left| \delta _{i}\right| )$.
Then, for given $e_{\alpha },a_{1},...,a_{n}$, the number of operators $%
V_{a_{n+1}}$ giving a nonzero contribution to $%
V_{a_{n+1}}V_{a_{n}}...V_{a_{1}}e_{\alpha }$. It does not exceed $k(\left|
\alpha \right| +C_{1}n),k=\left| Sub\right| $. Thus the number of nonzero
terms $V_{a_{n}}...V_{a_{1}}e_{\alpha }$ does not exceed 
\[
k^{n}\prod_{j=1}^{n}(\left| \alpha \right| +C_{1}j)=(kC_{1})^{n}\frac{(\frac{%
\left| \alpha \right| }{C_{1}}+n)!}{n!(\frac{\left| \alpha \right| }{C_{1}})!%
}\leq (kC_{1})^{n}n^{\frac{\left| \alpha \right| }{C_{1}}} 
\]
and the norm of each term is bounded by $\left( \max \lambda _{i}\right)
^{n} $. This gives convergence of the series for $\left| t\right| <t_{0}$
where $t_{0}$ does not depend on $\alpha $.

\subsection{$C^{\ast }$-algebra}

For each $N$ let $\mathcal{H}_{N}\subset \mathcal{H}$ be the finite
dimensional subspace generated by all $e_{\alpha }$ with $\left| \alpha
\right| \leq N$, let $P_{N}$ be the orthogonal projection onto $\mathcal{H}%
_{N}$. Let $\mathbf{A}_{N}$ be the $C^{\ast }$-algebra of all operators in $%
\mathcal{H}_{N}$. It is the $(\frac{r^{N+1}-1}{r-1}\times \frac{r^{N+1}-1}{%
r-1})$-matrix algebra if $r>1$. \ We can consider ''cut-off'' operators 
\[
a_{i,N}(j)=P_{N}a_{i}(j)P_{N} 
\]
as belonging to $\mathbf{A}_{N}$.

We have natural embeddings $\mathcal{H}_{N}\subset \mathcal{H}_{N+1}$ and we
define the embeddings $\phi _{N}:\mathbf{A}_{N}\rightarrow \mathbf{A}_{N+1}$
by: for $B\in \mathbf{A}_{N}$ we put $\phi _{N}(B)e_{\alpha }=Be_{\alpha }$
if $\left| \alpha \right| \leq N$ and $\phi _{N}(B)e_{\alpha }=0$ if $\left|
\alpha \right| =N+1$. The inductive limit $\cup _{N}\mathbf{A}_{N}=\mathbf{A}%
^{0}$ of the $C^{\ast }$-algebras $\mathbf{A}_{N}$ is called the local
algebra, its norm closure $\mathbf{A}$ is called the quasilocal algebra. It
does not fall however under the general definition of quasilocal algebras 
\cite{brro}, due to the absence of \ ''space structure''. There is no
identity element in this algebra (it can be appended if necessary, the
identity operator in $\mathcal{H}$), but there is an approximate identity, a
sequence $1(\mathbf{A}_{N})$ of unit matrices in $\mathbf{A}_{N}$. $\mathbf{A%
}$ is a hyperfinite $C^{\ast }$-algebra.

Note that the formal Hamiltonian $H$ defines the differentiation of the
local algebra. Denote 
\[
H_{N}=\sum_{i=1}^{\left| Sub\right| }\sum_{j=1}^{N}\lambda _{i}a_{i,N}(j) 
\]
Take some local $A$ and $N$ such that $A\in \mathbf{A}_{N}$. Define an
automorphism group of $\mathbf{A}_{N}$ as follows 
\[
\alpha _{t}^{(N)}(A)=\exp (iH_{N}t)A\exp (-iH_{N}t) 
\]

\begin{theorem}
There exists $t_{0}>0$ such that for any local $A$ and for each $t,\left|
t\right| <t_{0}$, there exists the norm limit 
\[
\lim_{N\rightarrow \infty }\alpha _{t}^{(N)}(A) 
\]
This defines a unique automorphism group of the quasilocal algebra.
\end{theorem}

Proof. Consider the Dyson-Schwinger series 
\[
A_{t}^{(N)}=A+\sum_{n=1}^{\infty }\frac{(it)^{n}}{n!}[%
H_{N},...,[H_{N},[H_{N},A]]...] 
\]
One can take $A=A_{\alpha \rho }$ where $A_{\alpha \rho }e_{\gamma }=\delta
_{\alpha \gamma }e_{\rho }$. Note that the commutator is the sum of
commutators 
\[
\lbrack a_{i_{n}}(j_{n}),...,[a_{i_{2}}(j_{2}),[a_{i_{1}}(j_{1}),A]],...] 
\]
multiplied by $\lambda _{i_{1}}...\lambda _{i_{n}}$. Nonzero commutators
should have the property that $j_{k}\leq l(A)+C_{1}(k-1),l(A)=\max (\left|
\alpha \right| ,\left| \rho \right| )$. The convergence proof is quite
similar to the previous convergence proof. If $N\rightarrow \infty $ then $%
A_{t}^{(N)}$ converge to 
\[
A_{t}=A+\sum_{n=1}^{\infty }\frac{(it)^{n}}{n!}[H,...,[H,[H,A]]...] 
\]
Each term of the latter series is well defined and the series converges for $%
t$ sufficiently small. The existence of the automorphism group can be proved
as in the Robinson theorem for quantum spin systems, see \cite{brro}.

\begin{remark}
Note that in the Robinson theorem for one-dimensional quantum spin systems
with finite interaction radius one can prove that the series converges for
all $t$, because the length of the cluster increases only at the boundary
(that is at two end points). For quantum grammars this is not the case.
\end{remark}

\subsection{KMS-states}

To define temperature states on $\mathbf{A}$ one could put for any local $A$
and large $N$ 
\[
<A>_{\beta }=\lim_{N\rightarrow \infty }<A>_{\beta ,N}=\lim_{N\rightarrow
\infty }Z_{N}^{-1}Tr_{N}\left[ A\exp (-\beta H_{N})\right] ,Z_{N}=Tr_{N}\exp
(-\beta H_{N}) 
\]
where $Tr_{N}$ means the trace in $\mathbf{A}_{N}$. However, this does not
always define a state. For example this gives zero for $\beta =0$, where $%
Z_{N}=\frac{r^{N+1}-1}{r-1}\rightarrow \infty ,r>1,Z_{N}=N+1\rightarrow
\infty ,r=1$, but $Tr_{N}A$ is bounded \ We shall prove it now in the
general case but only for small $\beta $.

\begin{lemma}
There exists $\beta _{0}>0$ \ such that for each local $A$ the limit 
\[
\lim_{N\rightarrow \infty }Tr\left[ A\exp (-\beta H_{N})\right] 
\]
exists and is analytic in $\beta $ for $\beta <\beta _{0}$.
\end{lemma}

Proof. Take again $A=A_{\alpha \rho }$. Then 
\[
Tr_{N}\left[ \exp (-\beta H_{N})A\right] =\sum_{k=0}^{\infty }\frac{(-\beta
)^{k}}{k!}Tr_{N}(H^{k}A)=\sum_{k=0}^{\infty }\frac{(-\beta )^{k}}{k!}%
\sum_{I_{k},J_{k}}Tr_{N}(a_{i_{k}}(j_{k})...a_{i_{1}}(j_{1})A) 
\]
where the sum is over all arrays $%
I_{k}=(i_{1},...,i_{k}),J_{k}=(j_{1},...,j_{k})$. The convergence proof is
the same as for the previous statements.

\begin{lemma}
If $\beta $ is small then $\log Z_{N}\sim cN$ with $c>1$. It follows that
the above limit $<A>_{\beta }$ is zero for all $A$.
\end{lemma}

Proof. One can write 
\[
Z_{N}=\sum_{\alpha :\left| \alpha \right| \leq N}z(\alpha ),z(\alpha
)=(e_{\alpha },\exp (-\beta H_{N})e_{\alpha }) 
\]
and 
\[
Z_{N}=\sum_{k=0}^{\infty }\sum_{I_{k},J_{k}}\sum_{\alpha :\left| \alpha
\right| \leq N}\frac{(-\beta )^{k}}{k!}z(\alpha ,(I_{k},J_{k})),z(\alpha
,(I_{k},J_{k}))=(e_{\alpha },a_{i_{k}}(j_{k})...a_{i_{1}}(j_{1})e_{\alpha }) 
\]
Consider some term of this expansion corresponding to some word $\alpha $ of
length $n$ and to some $(I_{k},J_{k})$. It is convenient to denote $\delta
(p)\rightarrow \gamma (p)$ the substitution on place $p$, corresponding to
the operator $b(p)=a_{i_{p}}(j_{p})$. The symbol $x_{i}$ of the word $\alpha
=x_{1}...x_{n}$ is called untouched for given $(I_{k},J_{k})$ if no $\delta
(p)$ contains it. Similarly, a symbol of the word $b(s)...b(1)e_{\alpha }$
is called untouched if no $\delta (l)$ with $s<l\leq k$ contains it.

We shall consider the lattices of partitions of words onto subwords. Let
some the word $\beta =\rho \gamma \kappa $ is obtained from the word $\alpha
=\rho \delta \kappa $ by the substitution $\delta \rightarrow \gamma $. Let
also a partition $G$ of $\alpha $ be given. We call a partition $G(\beta )$
of $\beta $ the partition induced by $G$ and the substitution $\delta
\rightarrow \gamma $ if following condition holds. If block $I$ of $G$
belongs to either $\rho $ or $\kappa $ then it is also a block of $G(\beta )$%
. The symbols of $\gamma $ form one block together with all symbols of the
blocks $I$ (not belonging to $\delta $) of $G$ intersecting with $\delta $.

Now let a partition $G$ of $\beta $ be given. We call a partition $G(\alpha
) $ of $\alpha $ the partition induced by $G$ and this substitution if
following condition holds. If block $I$ of $G$ belongs to either $\rho $ or $%
\kappa $ then it is also a block of $G(\alpha )$. The symbols of $\delta $
form one block together with all symbols (not belonging to $\gamma $) of the
blocks $I$ of $G$ intersecting with $\gamma $.

We define now inductively the set of partitions $G_{s}$ of partitions of the
words $\alpha _{s}=b(s)...b(1)\alpha ,s=0,1,...,n$, where $\alpha =\alpha
_{0},\alpha _{k}=\alpha $. $G_{0}$ is the partition of $\alpha _{0}$ onto $n$
separate symbols. $G_{s+1}$ is the partition of $\alpha _{s+1}$ induced by
the substitution $\delta (s)\rightarrow \gamma (s)$. Denote $G_{s,0}=G_{s}$.
If the partition $G_{s+1,p}$ of $\alpha _{s+1}$ is defined then $G_{s,p+1}$
is defined as the partition of $\alpha _{s}$ induced by the substitution $%
\delta (s)\rightarrow \gamma (s)$.

We need the partition $G_{0,k}$. Its blocks are at the same time the blocks
of the partition of the interval $\left[ 1,n\right] $. We call them clusters
with respect to $(\alpha ,I_{k},J_{k})$.

We call nonzero term of the expansion connected (for fixed $\alpha $ and $%
(I_{k},J_{k})$) if the partition $G_{0,k}$ consists of only one cluster.

Consider the contribution $c_{I}$ of some cluster $I$. It depends only on
its length $m$%
\[
c_{I}=c(m)=\sum_{\alpha :\left| \alpha \right| =m}\sum_{k=0}^{\infty
}\sum_{(I_{k},J_{k})}\frac{(-\beta )^{k}}{k!}(e_{\alpha
},a_{i_{k}}(j_{k})...a_{i_{1}}(j_{1})e_{\alpha }) 
\]
where the last sum is over all connected $(\alpha ,I_{k},J_{k})$. We have
the cluster expansion for $z(N)$ 
\[
z(N)\doteq \sum_{\alpha :\left| \alpha \right| =N}z(\alpha )=\sum
c_{I_{1}}...c_{I_{p}} 
\]
where the sum is over all partitions on consecutive intervals. To prove this
formula take the ordered array $\overrightarrow{m}=(m_{1},...,m_{p})$ of
positive integers such that $m_{1}+...+m_{p}=k$ and denote $%
\sum_{(I_{k},J_{k})}^{\overrightarrow{m}}$the sum over all $I_{k},J_{k}$
such that the numbers of substitutions touching the consecutive subwords $%
\alpha _{1},...,\alpha _{p}$, are correspondingly $m_{1},...,m_{p}$. Then 
\[
\sum_{(I_{k},J_{k})}^{\overrightarrow{m}}z(\alpha ,(I_{k},J_{k}))=\frac{k!}{%
m_{1}!...m_{p}!}\sum_{(I_{m_{1}},J_{m_{1}})}z(\alpha
_{1},(I_{m_{1}},J_{m_{1}}))...\sum_{(I_{m_{p}},J_{m_{p}})}z(\alpha
_{p},(I_{m_{p}},J_{m_{p}})) 
\]
We have also the cluster estimate 
\[
k(I)<c_{1}(C\beta )^{|I|} 
\]
It follows that $\log z(N)\sim cN$. Thus 
\[
\log Z_{N}\sim \log \left[ z(N)(1+\frac{z(N-1)}{z(N)}+...\right] \sim cN 
\]

\begin{remark}
Introduce the trivial substitutions $s\rightarrow s$ for each symbol $s\in S$
and denote $a(s;j)$ the correponding quantum sibstitutions. Let $P_{=N}$ be
the orthogonal projector onto the space $\mathcal{H}_{=N}=\mathcal{H}%
_{N}\ominus \mathcal{H}_{N-1}$. The cosmological term is defined as 
\[
\mu H^{0}=\mu \sum_{N=0}^{\infty }NP_{=N}=\mu \sum_{s\in
S}\sum_{j}a_{s}(j),\mu >0 
\]
Note that for Hamiltonians with the cosmological term 
\[
H_{N}+\mu H^{0} 
\]
the limiting state exists for $\mu $ sufficiently large as the partition
function is finite. It is natural to expect that there exists $\mu _{cr}=\mu
_{cr}(\beta )$ such that for $\mu <\mu _{cr}$ the limiting state does not
exist, but exists for $\mu >\mu _{cr}$. In most cases one can expect that
either $\log Z_{N}\sim cN$ or it is constant. It could be interesting to
know the cases when other possibilities occur. For example if $\beta =0$ and 
$S$ consists of one symbol only, then $\log Z_{N}\sim \log N$.
\end{remark}

\subsection{Classical space via renormalization}

Assume $\beta $ to be small as earlier and let us look at the ''support'' of 
$<>_{\beta ,N}$. More exactly, let $\mathbf{C}$ be the commutative $C^{\ast
} $-algebra, generated by multiplication (on bounded functions) operators in 
$\mathcal{H}=l_{2}(L)$. By restricting the state $<.>_{\beta ,N}$ on the $%
C^{\ast }$-subalgebra $\mathbf{C}_{N}=\mathbf{C\cap A}_{N}$, one gets the
measure $\mu _{\beta ,N}$ on the set of all words of length not exceeding $N$%
. One can show that as $N\rightarrow \infty $ the support of the measure $%
\mu _{\beta ,N}$ lies on the words of length of order $N$.

\subparagraph{Quantum Spin Systems}

We introduce some notation for quantum spin systems. Classical spin system
on $Z$ is a special probability measure on the set of configurations $S^{Z}$%
, that is functions on the ''space'' $Z$ with values in $S$. The space $Z$
has an additive group structure and acts on $S^{Z}$ as a group of
translations. The set of all words does not have such ''space structure''
but we shall show how the space (here it is $Z$), the quasilocal algebra on
this space, and a KMS state on this quasilocal algebra, can emerge from a
KMS-state on $\mathbf{A}$.

Consider classical spin configurations in a finite volume (that is the set $%
S^{\left[ -n,n\right] },\left[ -n,n\right] \subset Z$) as words of length $%
2n+1$. The Hilbert space for the corresponding quantum spin system is 
\[
\mathcal{K}_{2n+1}=\otimes _{i=-n}^{n}\mathcal{K}(i) 
\]
where $\mathcal{K}(i)$ is the $r$-dimensional Hilbert space with basis $%
e_{a},a=1,...,r$. Consider the\ $C^{\ast }$-algebra of linear operators in $%
\mathcal{K}_{2n+1}$: $\mathbf{L}_{2n+1}=\mathbf{W}_{-n}\otimes ...\otimes 
\mathbf{W}_{n}$, where $\mathbf{W}_{i}$ are $r\times r$-matrix algebras. The
quasilocal quantum spin algebra $\mathbf{L}$ is the norm \ closure of the
local algebra $\mathbf{L}^{0}=\cup \mathbf{L}_{2n+1}$.

Consider the Hilbert space $\mathcal{H}_{=n}=\mathcal{H}_{n}\ominus \mathcal{%
H}_{n-1}\subset \mathcal{H}$, generated by all $e_{\alpha },\left| \alpha
\right| =n$. Then $\mathcal{K}_{2n+1}$ can be naturally indentified with $%
\mathcal{H}_{=2n+1}$ by $\theta _{2n+1}:e_{\alpha }\in \mathcal{H}%
_{=2n+1}\rightarrow e_{a(1)}\otimes ...\otimes e_{a(2n+1)}\in \mathcal{K}%
_{2n+1}$ if $\alpha =a(1)...a(2n+1)$.

\begin{remark}
If $\left| \delta _{i}\right| =\left| \gamma _{i}\right| $ for all $i$ then
the subspaces $\mathcal{H}_{=n}$ are invariant and thus we get quantum spin
system Hamiltonians. In fact any quantum spin system Hamiltonians with
finite range interaction can be obtained as particular cases of the
Hamiltonians on quantum grammars by adjusting $\lambda _{i}$ appropriately.
\end{remark}

Consider the $C^{\ast }$-algebra $\mathbf{M}_{n}$ of linear operators in $%
\mathcal{H}_{=n}$. Consider the isomorphism $\chi _{2n+1}:\mathbf{L}%
_{2n+1}\rightarrow \mathbf{M}_{2n+1}$ induced by $\theta _{2n+1}$.

Consider the embeddings $\phi _{n}:\mathbf{M}_{2n+1}\rightarrow \mathbf{A}$,
given for $M\in \mathbf{M}_{2n+1}$ by 
\[
\phi (M)e_{\alpha }=Me_{\alpha },,\left| \alpha \right| =2n+1;\phi
(M)e_{\alpha }=0,\left| \alpha \right| \neq 2n+1 
\]
Consider some positive linear functional $\omega $ on $\mathbf{A}$. Then $%
\omega ^{\prime }=\omega \circ \phi _{2n+1}\circ \chi _{2n+1}$ is a positive
linear functional on $\mathbf{L}_{2n+1}$. By normalizing we get the state $%
<L>_{2n+1}=Z_{2n+1}^{-1}\omega ^{\prime }(L)$ on $\mathbf{L}%
_{2n+1},Z_{2n+1}=\omega ^{\prime }(1(\mathbf{L}_{2n+1}))$, where $1(\mathbf{L%
}_{2n+1})$ is the unit matrix in $\mathbf{L}_{2n+1}$. Consider the limiting
state on the quasilocal algebra $\mathbf{L}$ 
\[
<.>_{\beta ,Z}=\lim_{n\rightarrow \infty }<.>_{2n+1} 
\]
if the limit exists.

\begin{theorem}
If $\beta $ is small enough then the state $<.>_{\beta ,Z}$ exists and is a
KMS state on the quantum spin algebra $\mathbf{L}$.
\end{theorem}

Proof. Existence of the limiting state can be proven by cluster expansions.
We have two representations 
\[
Z_{2n+1}=\omega ^{\prime }(1(\mathbf{L}_{2n+1}))=\sum c_{I_{1}}...c_{I_{p}} 
\]
and for some $A\in \mathbf{L}_{2n+1}$ with support in $\left[ r,s\right]
\subset \left[ -n,n\right] $%
\[
\omega ^{\prime }(1(\mathbf{L}_{r+n})\otimes A\otimes 1(\mathbf{L}%
_{n-s}))=\sum_{-n\leq m\leq r,p\leq n-s}\omega ^{\prime }(1(\mathbf{L}%
_{m+n})))c_{m,l}(A)\omega (1(\mathbf{L}_{n-l})) 
\]
The first representation was proved earlier, the second can be proved
similarly. From these two representations the convergence to the limiting
state follows by standard techniques, see \cite{mm}. It also follows from
the cluster expansion that the limiting state is faithful, that is positive
for positive elements. Thus in the GNS representation $(\frak{M},\pi ,\Omega
)$ the cyclic vector $\Omega $ is separating. Then Tomita-Takesaki theory
defines a modular automorphism group of the von Neumann algebra and the
limiting state is the KMS with respect to the modular group of automorphisms.

\begin{remark}
From the cluster expansion one could get more. This KMS state is limit of
the states in finite volumes. Thus one could ask about the effective
Hamiltonian for the resulting quantum spin system. The effective hamiltonian 
$H_{eff}$ of this quantum spin system has non-finite multi-particle
potential, that is 
\[
H_{eff}=\sum_{i\in Z}\sum_{I}\tau ^{i}(\Phi _{I}) 
\]
where $\tau $ is the shift on $1$ in the spin quasilocal algebra on $Z$ and
the second sum is over all intervals $I$ containing $0$. Moreover, for all $%
I $ we have 
\[
\left| \Phi _{I}\right| \leq C\beta ^{\left| I\right| } 
\]
for some $C>0$.
\end{remark}

We will not prove the statement of this remark.

\begin{remark}
Note that the space structure can be obtained in different ways, using
different embeddings. For example, one can get a quantum spin system on $%
Z_{+}$ (and the space then will be $Z_{+}$), using the isomorphism 
\[
\mathcal{H}_{n}\rightarrow \otimes _{i=0}^{n-1}\mathcal{K}(i) 
\]
\end{remark}

\section{Quantum graph grammars}

\subsection{Definitions}

Labelled spin graph $\alpha =(G,s)$ is a graph $G$ with given set of
vertices $V=V(G)$ and a function $s:V\rightarrow S$, where $S$ is the spin
space. Further on we assume it to be finite. Two labelled graphs are said to
be equivalent (isomorphic) if they are isomorphic as graphs and the
isomorphism respect spins. Equivalence classes are called (unlabelled) spin
graphs. There are many other names for spin graphs: in physics spin graphs
are refered as spin networks, in computer science they are called also
marked graphs etc.

We remind definitions from \cite{m2}.

\begin{definition}
The substitution (production) $Sub=(\Gamma ,\Gamma ^{\prime },V_{0},\varphi
) $ is defined by two ''small'' spin graphs $\Gamma $ and $\Gamma ^{\prime }$%
, subset $V_{0}\subset V=V(\Gamma )$ and mapping $\varphi :V_{0}\rightarrow
V^{\prime }=V(\Gamma ^{\prime })$, either of $\Gamma $ and $\Gamma ^{\prime
} $ can be empty.

A transformation $T=T(Sub)$ of a spin graph $\alpha $, corresponding to a
given substitution $Sub$, is defined in the following way. Fix an
isomorphism $\psi :\Gamma \rightarrow \Gamma _{1}$ onto a spin subgraph $%
\Gamma _{1}$ of $\alpha $. Consider nonconnected union of $\alpha $ and $%
\Gamma ^{\prime }$, delete all links of $\Gamma _{1}$, delete all vertices
of $\psi (V)\setminus \psi (V_{0})$ together with all links incident to
them, identify each $\psi (v)\in \psi (V_{0})$ with $v^{\prime }=\varphi
(v)\in \Gamma ^{\prime }$. The function $s$ on $V(G)\setminus V(\Gamma _{1})$
is inherited from $\alpha $ and on $V(\Gamma ^{\prime })$ - from $\Gamma
^{\prime }$. We denote the resulting graph by $\alpha (Sub,\psi )$.

The graph grammar is a finite set of substitutions $Sub_{i},i=1,...,m$. We
call a graph grammar local if the $\Gamma $'s corresponding to all $Sub_{i}$
are connected. The language $L(\alpha _{0},\{Sub_{i}\})$ is the set of all
spin graphs which can be obtained from some initial spin graph $\alpha _{0}$
by applying transformations, correponding to $Sub_{i},i=1,...,m$, arbitrary
number of times in arbitrary order. More exactly, $\alpha _{0}\in $ $%
L(\alpha _{0},\{Sub_{i}\})$ and if $\alpha \in L(\alpha _{0},\{Sub_{i}\})$
then $T\alpha \in L(\alpha _{0},\{Sub_{i}\})$ for arbitrary $T=T(Sub_{i})$.
\end{definition}

The definition of a quantum graph grammar is similar to that of the quantum
grammar. Let $\frak{A}$ be a class of spin graphs, invariant with respect to
the substitutions of the given grammar, for example $\frak{A}=(\alpha
_{0},\{Sub_{i}\})$. Let $\mathcal{H}=\frak{A}$ be the Hilbert space with the
orthonormal basis $e_{\alpha }$ numerated by all spin graphs from $\frak{A}$%
: $(e_{\alpha },e_{\beta })=\delta _{\alpha \beta }$. \ For each spin graph $%
\alpha $ and each substitution $Sub_{i},i=1,...,m$, we enumerate somehow all
isomorphisms $\psi :\Gamma \rightarrow \Gamma _{1}$ as $\psi _{1},...,\psi
_{k(\Gamma )}$. Denote $a_{i}(j)$ the operator in $\mathcal{H}$ by $%
a_{i}(j)e_{\alpha }=e_{\alpha (Sub_{i},\psi _{j})}$ if $\psi _{j}$ exists,
that is if $j\leq k(\Gamma )$, and $0$ otherwise. Again we assume that
together with the substitution $\delta _{i}\rightarrow \gamma _{i}$ also its
''inverse'' substitution $\gamma _{i}\rightarrow \delta _{i}$ belongs to $%
Sub $ and the Hamiltonian is 
\[
H=\sum_{i=1}^{r}\sum_{j}\lambda _{i}a_{i}(j) 
\]
if $\lambda _{i}=\overline{\lambda }_{j}$ in case $\delta _{j}=\gamma
_{i},\gamma _{j}=\delta _{i}$. Note that the enumeration in $j$ has only
notational purpose, because the Hamiltonian is symmetric with respect to $j$.

\subsection{Examples}

We give here only two simplest examples.

\subsubsection{Mean field evolution on graphs}

There are no spins in this example. We consider 4 substitutions:

\begin{itemize}
\item  $Sub_{1}$ is defined by $\Gamma $ consisting of one vertex only, $%
V_{0}=V(\Gamma )$, $\Gamma ^{\prime }$ consisting of two vertices connected
by a link. The mapping $\phi $ just fixes one of these vertices. Then the
corresponding transformation consists of choosing a vertex $v$ of the graph $%
G$, appending a new vertex $v_{new}$ and connecting $v$ and $v_{new}$ by a
link.

\item  $Sub_{2}$ is the inverse sibstitution, that is we take a link having
at least one vertex of degree one and delete it.

\item  $Sub_{3}$ consists in appending a link between two chosen vertices.

\item  $Sub_{4}$ consists in just deleting a link.
\end{itemize}

This graph grammar is obviously nonlocal. The graphs can be non-connected
and we still denote them $\alpha $.

\subparagraph{Seladjointness}

We shall see now that a reasonable choice of the constants is 
\[
B=\sum_{j}(\lambda _{1}a_{1}(j)+\frac{1}{N}\lambda _{2}a_{3}(j)),H=B+B^{\ast
}=\lambda _{1}\sum_{j}(a_{1}(j)+a_{2}(j))+\frac{1}{N}\lambda
_{2}\sum_{j}(a_{3}(j)+a_{4}(j)) 
\]
where $N$ is the number of vertices in $\alpha $ and $\lambda _{1},\lambda
_{2}\geq 0$. Note that finite linear combinations of $e_{\alpha }$ are $%
C^{\infty }$-vectors for $L$. Denote this set by $D(L)$.

\begin{lemma}
$H$ is essentially selfadjoint on $D(L)$.
\end{lemma}

Proof. We shall prove that each vector from $D(L)$ is an analytic vector of $%
L$, that is 
\[
\sum_{k=0}^{\infty }\frac{\left\| H^{k}\phi \right\| }{k!}t^{k}<\infty 
\]
for some $t>0$. Note that

\[
\left\| B\right\| _{\mathcal{H}_{N}}=\lambda _{1}N+\frac{\lambda _{2}N^{2}}{N%
}=(\lambda _{1}+\lambda _{2})N 
\]
Here $\mathcal{H}_{N}$ is generated by $e_{\alpha },V(\alpha )\leq N$. Note
also that it is sufficient to take $\phi =e_{\alpha }$ and 
\[
He_{\alpha }=\sum_{\beta }e_{\beta } 
\]
where $V(\beta )\leq V(\alpha )+1$. The proof then is quite similar to the
one for the quantum grammars.

\subsubsection{Dual quantum evolution of two dimensional complexes}

Here we consider a more physical example corresponding to the pure quantum
gravity, where the quantum space has dimension $2$ and the time is classical
and has dimension $1$. Consider the set $\frak{T}$ \ of equivalence classes
of triangulations $T$ of closed oriented compact surfaces $S=S_{\rho }$ of
arbitrary genus $\rho =\rho (T)$, $N=N(T)$ is the number of triangles in $T$%
. A triangulation $T$ is defined by a pair $(G;\phi :G\rightarrow S)$ where $%
G$ is a graph and $\phi $ is its smooth embedding into $S$. Two
triangulations $T$ and $T^{\prime }$ are equivalent if there is a
homeomorphism $\phi :S\rightarrow S$ such that vertices of $G$ go to the
vertices of $G^{\prime }$, edges to edges, triangles to triangles.

It is more convenient to consider dual graphs $\Gamma =G^{\ast }$, the
vertices of $\Gamma $ correspond to the triangles of $G$. Two vertices of
the dual graph are connected by a link iff the corresponding triangles have
common edge. Thus each vertex of $\Gamma $ has degree $3$. Then the number $%
N $ of vertices of $\Gamma $ is even. The set $\frak{T}$ \ of equivalence
classes can be described equivalently in a purely combinatorial way in terms
of dual graphs, see \cite{wale}. Consider the set $\frak{G}$ of graphs $%
\Gamma $ with an additional structure. Each $\Gamma $ has $N=N(\Gamma )$
vertices, each of degree 3, $N$ is even. The additional structure on this
graph is defined as follows: for each vertex the cyclic order of its
edge-ends (legs) is fixed. It is not difficult to see that there is
one-to-one correspondence between $\frak{G}$ and $\frak{T}$.\ In one
direction it is trivial: take an embedding $\Gamma \rightarrow S$ and fix
orientation of $S$. Then choose say the clockwise order of the edge-ends in
each vertex.

Let $K$ be the set of edge-ends of $\Gamma $, it has then $3N$ elements.
There are two permutations on $K$: the first one $P$ consists of $V=N$
cycles of length $3$ and the second one $I$ consists of $E=\frac{3N}{2}$
cycles of length $2$. Then vertices of $\Gamma $ can be identified with
cycles of the permutation $P$, edges (links) of $\Gamma $ can be identified
with cycles of the permutation $I$, faces of $\Gamma $ can be identified
with cycles of the permutation $PI$.

We introduce the Hilbert space $\mathcal{H}=l_{2}(\frak{G})$ with the basis $%
e_{\Gamma }$. The evolution is defined as follows. For given $K=K(\Gamma )$
append $6$ new elements, thus $2$ cycles of length $3$. Choose $3$ edges $%
j_{1},j_{2},j_{3}$ (that is the cycles of length $2$) in $K$, cut them thus
getting $6$ other edge-ends, and reconnect $12$ edge-ends so that the
resulting graph were connected. The reconnection is done via some rule $\pi $
(depending on the set of $12$ edge-ends). This will give the linear operator
(quantum substitution) $a_{\pi }(j_{1},j_{2},j_{3})$.


The resulting Hamiltonian is defined on the subspace $\mathcal{H}_{N}$
generated by the graphs $\Gamma $ with $N$ vertices as follows 
\[
H=\frac{\lambda }{N^{2}}(B+B^{\ast }),B=\sum_{\pi
}\sum_{j_{1},j_{2},j_{3}}a_{\pi }(j_{1},j_{2},j_{3}) 
\]
where $j_{1},j_{2},j_{3}$ is an arbitrary unordered array of 3 links. The
adjoint term $B^{\ast }=\sum_{\pi }\sum_{v_{1},v_{2}}b_{\pi }^{\ast
}(v_{1},v_{2})$ describes the deletion of $2$ vertices, $v_{1},v_{2}$ is an
unordered array of $2$ vertices, $\pi $ describes how the the remaining
edge-ends are to be reconnected.

Note that the graphs here are not labelled. Thus one should be accurate with
the automorphisms. Remind that almost all $3$-regular graphs do not have
nontrivial automorphisms.

It leaves invariant the symmetrical subspace $\mathcal{H}_{symm}$ of the
Hilbert space, that is the space of functions of $L$ depending only on the
number of triangles $N=N(L)$ and on the genus $\rho =\rho (L)$. Note that $%
\mathcal{H}_{symm}$ is isomorphic to $Z_{+}^{2}$. We shall study the
spectral properties of this Hamiltonian in another paper.

\section{Comments}

\subsection{Context Free Grammars}

There are a lot of beautiful reviews on quantum computation now, see \cite
{tsi, ste, aha}. Quantum analogs of the standard computer science objects
are quantum Turing machines, quantum circuits, quantum automata, quantum
cellular automata etc. Our definition of a quantum grammar resembles
partially each of them. That of a quantum cellular automaton, but where the
lattice is a quantum object changing in time. Each term of the series
expansion constritutes a transformation defined by a quantum circuit. For
higher dimension we have a quantum analog of Kolmogorov-Uspenskij
algorithms, or a quantum analog of the graph grammars.

The operators $a_{i}(j)$ in our definition are homogeneous (do not depend on 
$j$) but it is easy to consider inhomogeneous analog, taking the set of
substitutions dependent of $j$. Then the time evolution of a quantum
grammars can also be looked at as a general quantum circuit.

In the computer science there are some peculiarities in the definition of
grammar. The alphabet $S$ is the union $S=T\cup W$ of two nonintersecting
alphabets: terminals $T$ and non-terminals (variables) $W$. The
substitutions (productions) $\alpha _{i}\rightarrow \beta _{i},i=1,...,m$,
are such that each $\alpha _{i}$ contains at least one symbol from $W$. The
quantum grammar is defined by the set of numbers $\lambda _{l}=\lambda
(\alpha _{i}\rightarrow \beta _{i})$, which are assumed to be real. In other
words by the linear operator 
\[
L=\lambda _{i}\sum a_{i}(j) 
\]
We prefer to use continuous time. If $L$ is not assumed to be symmetric then
one can define context free grammars, see \cite{mocr} (but there are no
nontrivial symmetric context free grammars) \ A context free grammar is one
where all $\alpha _{i}$ have length $1$, that is they are variables. Random
context free grammars were studied in \cite{m1} in a more general situation
when there is no subdivision of the alphabet.

For discrete time, which is assumed in \cite{mocr}, there are several ways
to define the evolution, i.e. the derivation. Discrete time analog of our
definition could be naturally given in a parallel form, that is all possible
substitutions are done for the word at the moment. This is easy to do for
context free grammars but not in more general cases. This is one of the
reasons to use continuous time what we do here.

Let $\mathcal{H}_{term}$ and $\mathcal{H}_{var}$ be the Hilbert subspaces of 
$\mathcal{H}$ defined by the corresponding parts of the alphabet $S$, and $%
P_{term},P_{var}$ are the orthogonal projections on these subspaces. The
derivation is the mapping 
\[
\lim_{t\rightarrow \infty }P_{term}e^{itL}P_{var}:\mathcal{H}%
_{var}\rightarrow \mathcal{H}_{term} 
\]
Otherwise speaking we start with some word from $W^{\ast }$ (the set of
words over $W$), even with the symbol from $W$, and stop each time when all
symbols in the resulting word are terminal. Note that $e^{itL}$ is the
identity on $\mathcal{H}_{term}$.

Existence of the dynamics can be proved quite similarly even in the
non-symmetric case.

\subsection{Spectrum}

We saw already that the lattice models of statistical physics and quantum
field theory constitute a particular case of the models on quantum lattices.
The most interesting question is the study of the spectrum of such models:
whether it has particles, scattering etc. We show below that the spectrum
have some new features even for the simplest models.

We already mentioned that the derivation for the grammar is decribed by the
operator $P\exp itH$, only large $t$ are interesting for us. If one knows
that $H$ is unitary equivalent to $H_{0}$ for some simple $H_{0}$, that is $%
H=UH_{0}U^{-1}$, then the operator $P\exp itH$ reduces to $P_{U}\exp
itH_{0},P_{U}=UPU^{-1}$. If $H$ describes something like interacting
infinite particle system then $H$ the $H^{0}$ can be the corresponding free
hamiltonian describing free quasiparticles. That is why spectral properties
of $H$ are related to the derivation in grammars.

Note that $e_{\emptyset }$ is a zero eigenvector of $H$. One could expect
that the rest of the spectrum of such operators should be similar to the
spectra of many particle systems. In particular one could expect that $H$ is
unitary equivalent to a free hamiltonian in a Fock space over some
one-particle subspaces. One could expect also that among these particles
some correspond to quanta of space and some - to quanta of matter fields.
Could one find an exact formulation of this statement ?

Detailed study of the spectrum of $H$ is necessary for this, and we shall do
in another paper, here we only give simplest examples. In the rest of the
paper we shall follow another idea: under some scaling we get a classical
space (here the lattice $Z$) and quantum spin system on it. This scaling
destroys thus the quantum character of space.

\begin{enumerate}
\item  (Quantum spin systems) We say that the hamiltonian $H$ is space (or
lattice) conserving if $\left| \delta _{i}\right| =\left| \gamma _{i}\right| 
$ for all $i$. Space conserving operators can be reduced to quantum spin
systems, as we shall see below. In this case there are only particles
corresponding to matter.

\item  (One-particle space) Let $r=1$, that is the alphabet consists of one
symbol $a$. Consider two substitutions $1:a\rightarrow aa,2:aa\rightarrow
a,\lambda _{1}=\lambda _{2}=\lambda $, and the Hamiltonian 
\[
H=\lambda \sum_{j=1}^{\infty }(a_{1}(j)+a_{2}(j)) 
\]
with real $\lambda $. Then the Hilbert space $\mathcal{H}$ is isomorphic to $%
l_{2}(Z_{+})$, because the word $aa...a$ can be identified with its length
minus $1$. The Hamiltonian is unitary equivalent to Jacobi matrix 
\[
(Hf)(n)=\lambda (n-1)f(n-1)+\lambda nf(n+1) 
\]
We shall call this operator a one-particle operator, this ''particle'' is
natural to associate with a space quanta. Here the space evolution is the
simplest one (due to one dimension): expansion and compression (in each
point). We shall find its spectrum in another paper.

\item  Consider $S=\left\{ a,w\right\} $ and the following substitutions 
\[
a\rightarrow aa,aa\rightarrow a,aw\rightarrow wa,wa\rightarrow aw 
\]
The subspace $\mathcal{H}_{1}$ generated by words with exactly one symbol $w$
is invariant. This hamiltonian can be interpreted as a mixture of the
previous pure space hamiltomian $H_{1}$ and the discrete laplacian $H_{2}$,
the free nonrelativistic one-dimensional Schroedinger operator in $l_{2}$ on
a finite set. Similarly, the subspace $\mathcal{H}_{2}$ with exactly two
symbols $w$ corresponds to two matter particles.

\item  (Noncommutative Fock space) Let $S=\left\{ a,b\right\} $ and consider
4 substitutions 
\[
1:a\rightarrow aa,2:aa\rightarrow a,3:b\rightarrow bb,4:bb\rightarrow b 
\]
Here there are two invariant one-particle spaces $\mathcal{H}_{a},\mathcal{H}%
_{b}$. For example, $\mathcal{H}_{a}$ is generated by words $%
a,a^{2}=aa,...,a^{n},...$. There are two invariant two-particle spaces $%
\mathcal{H}_{ab}=\mathcal{H}_{a}\otimes \mathcal{H}_{b},\mathcal{H}_{ba}=%
\mathcal{H}_{b}\otimes \mathcal{H}_{a}$. For example, $\mathcal{H}_{ab}$ is
generated by words $a^{k}b^{l},k,l>0$. In general for each even $n$ there
two invariant $2n$-particle spaces $\mathcal{H}_{(ab)_{n}}$, generated by
words $a^{k_{1}}b^{l_{1}}...a^{k_{n}}b^{l_{n}}$, and $\mathcal{H}_{(ba)_{n}}$%
. Similarly for odd $n$ there are two invariant $(2n+1)$-particle spaces $%
\mathcal{H}_{b(ab)^{n}},\mathcal{H}_{(ab)^{n}a}$. For arbitrary $r>2$ with
substitutions $i\rightarrow ii,ii\rightarrow i$ for each $i=1,2,...,r$ we
have $r(r-1)^{n-1}$ $n$-particle spaces. Thus, the standard spectrum of
tensor products has the corresponding multiplicities. This example supports
the name ''one-particle'' in the first example, because we get here a Fock
space OVER these two one-particle spaces.
\end{enumerate}

This shows a rich structure of the introduced Hamiltonians.

\subsection{Short overview of evolution types}

We give here a very short overview of other papers where related dynamics of
discrete structures were considered. Note that most papers have more
geometric and algebraic aspect than analytic one. In our paper we considered
mainly analytic problems.

\paragraph{ Deterministic evolution}

Deterministic evolution of words is one of the main subjects of the computer
science. In computer science marked graphs and their deterministic evolution
were known since Kolmogorov-Uspenskij paper \cite{kous}. Now there is a
large field in computer science, which studies graph grammars - local
dynamics of the marked graphs. In \cite{req2} deterministic evolution of
classical spin systems on graphs is defined. The basic graph is fixed or
taken randomly via random graph theory procedure, that is for fixed set of
vertices each bond is drawn independently with some probability $0<p<1$.

\paragraph{Markov processes}

Random grammars were considered earlier in computer science context, as
Markov processes. But questions related to the thermodynamic limit appeared
only in \cite{m1, m2}.

\paragraph{Unitary evolution and causal structure}

Such evolution is the main object in quantum computing in the computer
science context and in quantum gravity in a physical context. The latter
considers spin graph as a quantum object, thus one deals with the wave
function on the set of all possible spin graphs. The square of the wave
function defines a probability \ distribution on spin graphs.

Spin networks (graphs with spins, half-integers, living on the links, and
some operators in the vertices) were introduced in physics by R. Penrose 
\cite{penrose}. In physics now there are many variants of the quantum
evolution, discussed in \cite{marko, boris, kasm, smolin2, thie1, masm1,
masm2, baez1} as well as in earlier papers, see \cite{dapi}. Wider
generalization are discrete complexes with a causal structure which could
model Lorentzian structure on manifolds.

\paragraph{Completely positive semigroups}

This case is interesting due to inevitable noise coming from the environment
of the quantum system. See discussion of these problems in \cite{aha}.

\paragraph{Nonlinear Markov processes}

There can be transformations of probability measures on spin complexes,
which cannot be reduced to a Markov process (that is they are not given by
random point transformations) and they are not of quantum mechanical nature.
Examples of such dynamics one can find in two-dimensional quantum gravity,
see \cite{ma3}.

\end{document}